# Karl Popper's Forgotten Role in the Quantum Debate at the Edge between Philosophy and Physics in 1950s and 1960s

Flavio Del Santo

Institute for Quantum Optics and Quantum Information – Vienna (IQOQI), Austrian Academy of Sciences Basic Research Community for Physics (BRCP)

**Abstract.**
It is not generally well known that the philosopher Karl Popper has been one of the foremost critics of the orthodox interpretation of quantum physics for about six decades. This paper reconstructs in detail most of Popper's activities on foundations of quantum mechanics (FQM) in the period of 1950s and 1960s, when his involvement in the community of quantum physicists became extensive and quite influential. Thanks to unpublished documents and correspondence, it is now possible to shed new light on Popper's central –though neglected– role in this "thought collective" of physicists concerned with FQM, and on the intellectual relationships that Popper established in this context with some of the protagonists of the debate over quantum physics (such as David Bohm, Alfed Landé and Henry Margenau, among many others). Foundations of quantum mechanics represented in those years also the initial ground for the embittering controversy between Popper and perhaps his most notable former student, Paul Feyerabend. I present here novel elements to further understand the origin of their troubled relationship.

## 1. Introduction: motivations, methods and goals.

Karl R. Popper (1902-1994) has been one of the most preeminent philosophers of science of the twentieth century. He is generally known for his views on scientific method, having introduced falsifiability as the demarcation criterion between scientific and non-scientific theories, and as a matter of fact still today "many scientists subscribe to some version of simplified Popperianism" (Kragh 2013).

However, it has been essentially neglected by contemporary historiography that Popper contributed to the scientific discourse in at least another manner: for about six decades he was in the front line of the controversy over quantum physics, side by side with some of the most eminent physicists of his time. Indeed, since as early as 1934, Popper was among the few opponents of the leading anti-realistic and instrumentalist *Copenhagen interpretation* of quantum mechanics (CIQM).[1]

Pushing forward a line of research introduced in a recent work (Del Santo 2017), this paper, without any pretense of being exhaustive, aims at proving a historical account of Popper's activities in the foundations of quantum mechanics (FQM) between 1950s and 1960s, with a focus on the intellectual relationships that Popper established with many influential physicists.

In this respect, I deem it advantageous to (loosely) base my reconstruction of Popper's work on Fleck's idea of *Denkkollektiv*. In 1935, Ludwig Fleck proposed the concept of *Denkkollektiv* (*thought collective*), as the "community of persons mutually exchanging ideas or maintaining intellectual interaction" (Fleck 1935). Although I am aware that it can be pointless and even pernicious to try to trace a sharp line between disciplines, I wish to point out the existence of an underlying structure of social nature in the intellectual communities, that strongly conditions the development of particular directions of research, introducing a specific methodology, terminology, "mood" and "thought style" (Fleck 1935).

---

[1] The authorship of the CIQM, and whether it is meaningful to speak of a Copenhagen interpretation at all, has been often questioned. Interestingly, Don Howard, in his insightful paper on the history of CIQM, even points out that "it was Popper, an old critic of quantum orthodoxy […], who did more than anyone else, starting in the late 1950s, to cement in the popular mind the idea that Bohr and Heisenberg were together committed to a subjectivist interpretation of quantum mechanics." (Howard 2004). In what follows I still maintain the historical definition of CIQM as an anti-realistic and instrumentalist interpretation of the quantum formalism, which shares some fundamental principles (e.g. an observer-dependant collapse of the wave function, the Born rule, the wave-particle duality, the correspondence principle, etc.).

In fact, Popper has been, in certain periods, an important link that joined the *Denkkollektiv* of philosophers of science – of which he was one of the most renowned members – with the *Denkkollektiv* of those illustrious outsider physicists concerned with reclaiming realism in quantum mechanics (QM).[2] These physicists were the "dissidents" who – having a common ground in a realistic interpretation of quantum physics – challenged orthodoxy (see Freire 2014). After the early critiques put forward by Einstein, Schrödinger, de Broglie, etc., the instrumentalist CIQM became widely accepted, also due to the pragmatic paradigm that Cold War demands had enforced in science. In the post-war period, research on FQM was kept alive – although with drastic conceptual and ideological differences – primarily by physicists the likes of D. Bohm, J. P. Vigier, H. Everett, B. DeWitt and J. Bell and slowly experienced a revival in the seventies thanks to other groups, especially in Italy, initiated by F. Selleri (see Baracca, Bergia, and Del Santo 2016), and in a peculiar way in the US too (see Kaiser 2011).

In the period here investigated, Popper's role in the debate regarding FQM transformed tremendously. I will show that from 1948 – when he came back to quantum mechanics – up until late 1960s, Popper gave innovative contributions to FQM. However, his activities were conducted mainly in the context of a philosophical *Denkkollektiv*. Indeed, Popper's rare publications on the subject matter until late sixties were diffused mainly in the distribution channels of philosophers (conferences, specialized journals, etc.). At the same time, Popper's regular interlocutors amongst professional physicists interested in the FQM were sparse, and essentially restricted to Alfred Landé, David Bohm and Hermann Bondi (who was an illustrious cosmologist, but never contributed to FQM).

From 1966 onwards, however, thanks to new acquaintanceships, such as Mario Bunge and Wolfgang Yourgrau and their editorial initiatives, Popper found room in the Physics *Denkkollektiv*. It is the case of Popper's first paper deliberately though for an audience of physicists, *Quantum Mechanics without the Observer* (Popper 1967), of which the contents represent the first organic exposition of his own realistic interpretation of quantum mechanics, and of his mature arguments against CIQM (see Section 3.2). Section 3.3 deals with the initial resonance that this paper had in the *Denkkollectiv* of physicists. In this regard, it is also very interesting to notice that one of the most severe criticisms of (Popper 1967) was levelled by his former pupil Paul Feyerabend. By means of their exchange of correspondence, Section 3.4 is dedicated to reconstructing their diatribe on QM, relating this episode to the discourse on the complex relationship between Popper and Feyerabend (see e.g. Collodel 2016).

It ought to be remarked that, at the same time, Popper started a crusade against the so-called Logic of Quantum Mechanics, an axiomatic approach to quantum theory in the spirit of CIQM, which, initiated by G. Birkhoff and J. von Neumann in 1936, was experiencing a revival in the 1960s. Indeed, Popper published a critical paper in the influential journal *Nature* (Popper 1968) and, consequently, he entered a harsh debate with a number of distinguished physicists.[3]

Although these activities had a certain resonance and undoubtedly an influence on the physics *Denkkollectiv*, the vast majority of this information remained confined to private meetings and correspondences, and never appeared in print thus far.[4] Consequently, not only historiography has hardly devoted any attention to the topic at issue, but the relevance of Popper's large involvement in FQM was often not understood even by most of his contemporaries. This is crystallized by the fact that in Popper's volume of "The library of Living Philosophers" (Schilpp 1974) – supposed to outline the major intellectual contributions of the foremost living philosophers– devoted essentially no attention to Popper's contributions in the field of quantum foundations. Such a lack of consideration was severely criticized by Popper's pupil W. Bartley,[5] who published an instructive account of Popper's critical positions about quantum physics (Bartley 1978), wherein, perhaps for the first time, it was emphatically highlighted how much Popper's contributions to physics had been always overlooked:

> *In fact, the physicists have for the most part simply ignored Popper.*
> Among the many highly distinguished contributors to the Schilpp volume there is, for example, only one physicist of even moderate distinction, Henry Margenau […]. There could be no more serious editorial defect in a Schilpp volume. It is as if the Einstein volume of the series were published without critical reference to relativity theory; or the Russell volume, without reference to mathematical logic. Where is Paul Feyerabend, who published a highly critical study of Popper's contributions to physics in 1968? Where is J. S. Bell, John F. Clauser, or John A. Wheeler? They certainly know of Popper's work […]. Were Werner Heisenberg, Kurt Gödel, Louis de Broglie, or Eugene P. Wigner asked to contribute? For that matter, where are Jean-Pierre Vigier, Alfred Landé, and David Bohm, who tend to agree with Popper's critique of quantum mechanics […]?
> *The lack of a sustained critical interaction between Popper and the majority of physicists is a loss not only to theory but also to culture.* (Bartley 1978, pp. 676-677. The emphasis are mine)

---

[2] An alternative, and perhaps valid description could be that of identifying a unique transdisciplinary *Denkkollektiv* made of both philosophers and physicists, whose central aim was to disprove the Copenhagen Interpretation and vindicate scientific realism. However, I deem it important to distinguish between these two collectives, because they surely did not always share the same scopes and the specific terminology, and only seldom had a mutual interaction and influence (especially in the case of professional collaborations and academic publications).

[3] It goes beyond the scope of the present paper to expound Popper's critique to the Logic of Quantum Mechanics that turns out to be a rather technical matter and alone would require a manuscript possibly more voluminous that the present one. Partial results on these specific matters have been presented by the author at the 37th National Congress of the Italian Society for the History of Physics and Astronomy in Bari (Flavio Del Santo, September 28, 2017. https://www.youtube.com/watch?v=2S_lWfGAyNA) and a research paper on the subject is in preparation. Proceedings of this congress are also in press.

[4] The present paper is vastly based on original documents (Popper's private correspondence, unpublished papers and notes) retrieved in the course of a research conducted at the Popper Archive (herein after PA in the in-text citations) of the Alpen Adria University in Klagenfurt (Austria).

### 2. Popper and QM in the fifties: the philosophical *Denkkollektiv*

Discouraged by an unfortunate experience of a flawed *Gedankenexperiment* he had proposed in his *Logik der Forschung* (Popper 1934), Popper set his interest in QM aside for more than a decade (see e.g. Del Santo 2017 and references thereof). It was only in 1948 that he came back to the problem of quantum mechanics, thanks to the discussions with the Austrian physicist Arthur March (see Popper 1976, p. 106). On November 15$^{th}$, 1948, Popper gave a talk at the British Society for the History of Science on "Indeterminism in Quantum Physics and in Classical Physics" and presented it in his lectures at Harvard University and again in 1950 in Princeton, in the presence of Einstein and Bohr (see Jammer 1991); Popper then expanded the subject of his talks in a paper (Popper 1951). Therein, Popper supports an indeterministic position, considering classical (Newton's and Maxwell's) theories only *prima facie* deterministic, yet showing that they can be "indeterministic in perhaps an even more fundamental sense than the one usually ascribed to the indeterminism in quantum physics (in so far as the unpredictability of the events […] is not mitigated by the predictability of their frequencies)" (Popper 1951). However, despite its title, this line of research cannot, strictly speaking, be considered part of Popper's activity on FQM. In fact, quantum physics is in this context used by Popper as a mere framework to compare different forms of indeterminism, but it is never the main object of investigation. Moreover, Popper's indeterminism found its origins also in ideological reasons against historicism, and, as such, scientific determinism is merely seen as a stronger form (i.e. easily disprovable) of metaphysical determinism, which remains the actual target of Popper's critique. Therefore, this argument, whatever remarkable, is to be actually considered part of the general philosophical viewpoint of Popper and its scope is confined to philosophy proper.

A breakthrough in Popper's conception of FQM came around 1953, when he developed a new interpretation of probability which he named *propensity interpretation*.[6] In fact, Popper realized that the *frequency interpretation* of probability, which he upheld till then, could not satisfactorily justify a realistic account of quantum interference and in particular of the two-slit experiment. This convinced Popper "that probabilities must be 'physically real' – that they must be physical propensities, abstract relational properties of the physical situation" (Popper 1959). In April 1957, at a symposium in Bristol,[7] the propensity interpretation was publicly presented for the first time, and some works appeared in print soon after (Popper 1957; 1959). However, in both cases the connection with quantum physics was superficial and their diffusion in the physics *Denkkollektiv* negligible: whilst the first paper was an initial attempt that only appeared in the conference proceedings, the latter was in fact a formal paper on probability theory and its foundations. Although Popper clearly states that the first motivation (both "in time and importance") for which he gave up the frequency interpretation "was connected with the problem of the interpretation of quantum theory" (Popper 1959), he then explicitly points out that his paper is devoted only to discussing some flaws he found in his own treatment of probability, and QM remains drastically marginal. Moreover, the second paper was published in *The British Journal for the Philosophy of Science*, which further reduced the possibility that any physicist would notice Popper's results at that time. It was only with his paper "Quantum Mechanics without the Observer" (Popper 1967) that Popper fully formulated the propensity interpretation in the context of quantum physics (see Section 3.2).

---

[5] Bartley was the pupil of Popper who above all the others devoted his interest towards Popper's philosophy of QM. In 1982 he edited Popper's *Postscript to the Logic of Scientific Discovery*, whose third and last volume (Popper 1982) is the most comprehensive account of Popper's work on quantum mechanics. Therein, Popper also firstly published his EPR-like thought experiment (see Del Santo 2017) and reprinted his "Quantum Mechanics without the Observer" (Popper 1967).

[6] For the historical development of propensities see (Popper 1982), p. 95, footnote 98 and (Jammer 1974), p. 448 ff.

Between 1951 and 1956, Popper had put a great deal of effort to update and systematize his critical ideas about quantum theory that should have appeared in an extension of his *Logic of Scientific Discovery* – the English edition of (Popper 1934), published in 1959. However, due to accidental circumstances,[8] this work was published only in 1982 in a volume entitled *Quantum Theory and the Schism in Physics* (Popper 1982). Nevertheless, the preprint of this book circulated among Popper's closest students and colleagues at least since 1957, but most likely had absolutely no influence on the *Denkkollektiv* of physicists.

For what concerns the physicists, in fact, up until the whole decade of 1950s, Popper maintained good relationships with a few eminent physicists interested in fundamental and philosophical problems. A physicist with whom Popper had a certain affinity and established a long-lasting relationship was Alfred Landé (1888-1976). After having received a doctorate under A. Sommerfeld and served as an assistant to D. Hilbert, Landé gave pivotal contributions to atomic and quantum physics in the 1920s (as the explanation of the anomalous Zeeman effect, whose solution involves the so-called Landé factor). However, although Landé had regular contacts with Popper since at least 1953,[9] he only started quoting Popper's works rather late, in his book *From Dualism to Unity in Quantum Mechanics* (Landé 1960). Nonetheless, therein the propensity interpretation of probability is never mentioned, suggesting that Landé was still unaware of it. Popper's influence on Landé's works became manifest only throughout the sixties, as testified by Landé himself when recalling that "in the new manuscript of *From Dualism to Unity* [Popper was] quoted at least twenty times as chief witness." (Letter to Popper on March 11th, 1963. PA 318/18). This increasing appreciation towards Popper is reflected in a letter by Landé, wherein he affirmed: "I am your most staunch adherent among the physicists" (March 11th, 1963. PA 318/18).

Another physicist who established a close relationship with Popper and was to become one of the most influential intermediaries between Popper and the physics community, was David Bohm (1917-1992). Bohm, "one of the most significant theoretical physicists of the 20th century" (Peat 1997, p.

---

[7] In a very recent and interesting publication, it has been claimed that this Ninth Symposium of the Colston Research Society was "the first major event after World War II" about foundations of quantum mechanics (Kožnjak 2017). Some of the most prominent physicists in the field, including Bohm who had just become Professor at the University of Bristol, participated to the event together with distinguished philosophers of science. It is however important to notice that Popper did not attend the symposium and sent a paper read by Feyerabend instead. Popper thus missed a momentous occasion for a direct confrontation with the *Denkkollektiv* of physicists, which could have speeded up his active involvement in quantum physics. It is also worth mentioning that at that time Popper was not yet aware of J. P. Vigier's work on the FQM. In fact, in a letter to Popper Feyerabend explained that "Vigier [was] a pupil of de Broglie and trying to work out an interpretation of quantum-mechanics which is similar to Bohm's" (January 14th, 1957. PA 294/19). Vigier has been perhaps the single greatest influence on Popper's late contributions to QM (see Del Santo 2017).

[8] For a detailed description of the complex editorial process that the *Postscript* underwent, see (Popper 1982, editor's preface) and (Popper 1976, p. 172 ff.).

[9] The first documented correspondence is a letter from Popper to Landé, on June 21st, 1953. (PA 548/8)

316), gave his momentous contribution to FQM, reviving the old idea of *hidden variables* (initiated by Einstein and L. de Broglie), and providing the first deterministic model able to reproduce all the statistical results of quantum theory (Bohm 1952). Bohm became victim of the hateful persecution of McCarthyism, which forced him to leave US and eventually led him, from 1957, to settle in England. There, as a professor of physics first in Bristol and then at the Birkbeck College of the University of London, he soon made Popper's acquaintance.[10] Bohm and Popper had an intense intellectual relationship and whilst Bohm helped Popper a great deal in getting closer to the physics *Denkkollektiv*, Bohm largely profited by his interactions with Popper for what concerns philosophical issues. In this respect, physicist Roger Penrose remarked: "There can be few physicists who have delved into the philosophical implications of their subject as has David Bohm" (see Hiley 1997). It should be noticed, however, that Bohm and Popper's relationship was not completely free from conflicts.[11] Their friendship slowly deteriorated and, as a matter of fact, Bohm did not participate to Popper's critical activities on QM in the 1980s, which were stimulated by J.-P. Vigier instead (see Del Santo 2017).

A few words deserve to be said also about Hermann Bondi (1919-2005), who started a regular interaction with Popper around mid-1950s,[12] resulting in an everlasting friendship. Bondi was a distinguished cosmologist – together with F. Hoyle and T. Gold he put forward the popular "steady state theory" as an alternative to the Big Bang – but he became increasingly interested in the FQM. More than likely, it was due to Popper that Bondi developed a criticism towards the CIQM, and he supported Popper in many of his initiatives.

From 1947-48, Popper was in regular touch also with one of the founding fathers of QM, the Austrian Nobel laureate Erwin Schrödinger; however, their "relation had been somewhat stormy" (Popper 1976, p. 156). At any rate, the focus of their numerous discussions on philosophy of science was almost never quantum theory and thus I will not discuss this further.

It is very important to stress that, up until 1960s, Popper was for these physicists an interlocutor for what concerns only the genuinely philosophical matter and they very rarely quoted Popper in their publications on FQM. It is thus not surprising to notice that both Bohm and Landé, for instance, asked for Popper's advice and help to publish in distinguished journals on philosophy of science,[13] legitimizing Popper's role as a reference point (in the *Denkkollektiv* of philosophers) for the physicists with philosophical proclivities. On the other hand, there are no evidence that Popper ever tried to publish in physics journals or to spread his ideas among the physicists' collectives, until the late 1960s.

### 3. Popper and QM in the late sixties: the physics *Denkkollektiv*

#### 3.1 The road towards physics

In the meantime, the Argentine physicist and philosopher of science Mario Bunge (1919-) became interested in Popper's critical ideas about Copenhagen interpretation, after writing the paper "Strife about Complementarity" (Bunge 1955).[14] Bunge was to become for Popper a major connection to the *Denkkollektiv* of physicists: in 1961 he organized a *Festschrift* for Popper's sixtieth birthday (Bunge 1964).[15] Although Popper was "very happy, especially, about the scientists in the list",[16] it ought to be remarked that among the 22 contributors only Bohm and Yourgrau were quantum physicists (and Bohm's contribution was not even about QM). Indeed, Popper suggested to add Viktor Weisskopf and Landé, neither of whom contributed. Moreover, Bunge invited Popper to serve as co-editor of a new series of books, *Fundamenta Scientiae*. Popper accepted, and he was, besides Patrick Suppes, the only non-physicist among the co-editors. Most interestingly, Bunge explicitly expressed that the motivation of this invitation was also "to 'sell' [Popper] to the scientific public interested in [philosophical] questions".[17]

---

[10] The first available letter between Popper and Bohm dates September 5th, 1959. However, an envelope in Popper's possession, containing Bohm's first paper on hidden variable, is addressed to Bohm's institute in Brazil, when we stayed only until 1955 (PA 278/2).

[11] A particular divisive issue was triggered when Bohm became closer to the Indian philosopher and teacher Jiddu Krishnamurti. A biographer of Bohm reports that "Popper was sympathetic to Bohm's ideas until the topic of Krishnamurti was raised. To Popper, the Indian teachings smacked of totalitarism" (Peat 1997, p. 218). Moreover, still according to Peat, it seems that in the mid-1960s Popper somehow forced Bohm to erase a sentence from his contribution to a *Festschrift* in Popper's honor. "Bohm acquiesced, but he lost his earlier closeness to the philosopher; they no longer met on a regular basis" (Peat 1997, p. 218). Analyzing Popper and Bohm's correspondence, it seems that actually at least until the end of 1960s they kept in good touch, but as a matter of fact their relationship was undermined over time.

[12] The first letter between Popper and Bondi dates January 1st, 1956 (Bondi to Popper, PA, 530/10). Although previous interactions between the two are mentioned in that letter, it is possible to understand that their relationship was still quite formal, and probably recently established.

[13] Letters from Bohm to Popper on December 15th, 1960 (PA 278/2) and from Landé to Popper on January 18th, 1955 (PA 318/18).

[14] I am thankful to Prof. Mario Bunge for having shared his witness on this in a private communication (Mario Bunge, email to the author, May 7, 2017). The first letter between Bunge and Popper dates August 6th, 1957 (531/20).

But the real opportunity for Popper to connect with quantum physicists, was again provided by Bunge when he organized in Oberwolfach (Germany) a Symposium on the Foundations of Physics; Popper was invited together with 30 between physicists and philosophers.[18] He willingly accepted, maintaining: "I too have been thinking much (as always) about quantum theory; and I do no longer believe in everything I wrote about it in 1934 (in my *Logik der Forschung*)". It so appears that time was ripe for Popper to let the seeds that he had planted during the previous decade grow into a coherent exposition of his views about QM. Indeed, Popper immediately started writing the paper focused on his own objective-realistic interpretation of QM against CIQM. Although he eventually could not attend the conference in Oberwolfach for health reasons,[19] Bunge proposed to collect the realist-oriented essays prepared for that occasion in a collective volume for the new series of book that he was editing.[20] Therein, Popper's paper eventually was published under the title "Quantum Mechanics without the Observer" (Popper 1967), and was to have a remarkable impact on Popper's role in the physics *Denkkollectiv*.

At the end of 1950s,[21] Popper established a sincere friendship with another physicist who turned to philosophy of science, Wolfgang Yourgrau (1908-1979), more than likely thanks to their "mutual friend A. Landé" (Letter from Yourgrau to Popper on April 5th, 1963; PA 364/11). Yourgrau, who had been a student of Einstein in Berlin, became a passionate supporter of realism, fighting against the CIQM. In May 1962, when asked to edit 13 papers on the "world's leading men who contributed to the philosophy of science", Yourgrau proposed to include Popper as the only non-scientist.[22] Prolific writer and editor, during the 1960s, Yourgrau promoted Popper's critical activities on FQM with increasing vigor: he was one of the very few physicists who fully recognized the relevance of Popper's contribution to FQM, and for years he tried to persuade Schilpp to include them into Popper's volume of the *Library of Living Philosophers* (Schilpp 1974; see Section 1). In this regard, Yourgrau wrote to Popper:

> Margenau as well as Landé are of the opinion that I am better qualified to judge your contribution to theoretical physics than any other scholar they know. All of this will not influence Schilpp, I am afraid. You and I are personal friends, not merely colleagues, and that is for him a defect. (July 2nd, 1966; PA 364/11)

In June 1966, Yourgrau was asked to edit a *Festschrift* in honor of Landé and insistently persuaded Popper to contribute something on quantum theory, amongst physicists the likes of L. de Broglie, M. Born, E. Wigner, L. Rosenfeld, A. Kastler, F. Bopp.[23] Popper contributed with a paper

---

[15] Letter from Bunge to Karl and Hennie Popper on March 13th, 1961. (PA 280/25).

[16] Letter from Hennie Popper to Bunge on March 19th, 1961 (PA 280/25).

[17] Letters from Bunge to Popper on January 20th and 27th, 1966. (PA 280/26).

[18] Letter from Bunge to Popper on October 21st, 1965 (PA 280/25).

[19] Letters from Popper to Bunge on June 18th, 1966. (PA 280/26).

[20] Bunge announced this to Popper in a letter dated February 3rd, 1966. The Book was eventually printed under the title *Quantum Theory and Reality* (PA 280/26). The book collected contribution by outstanding physicists the likes of P. Bergmann, Vigier and Margenau.

[21] In a letter dated June 14th, 1955, Popper writes that he has "come across Yourgrau, not personally, but on Appointment Committees and [he] was not too favourably impressed" (PA 386/6). However, already in May 1959 Popper helped Yourgrau a great deal in finding an academic position, sending tens of letters of support for him (PA 394/18 and PA 364/11).

[22] In fact, clearly this collection was more physically oriented than philosophically: in the list stand the names of H. Helmholtz, E. Mach, L. de Broglie, M. Plank, A. Einstein, N. Bohr, E. Schrödinger. Yourgrau writes this to Popper hand his wife on May 23rd, 1962 (PA 364/11).

[23] In this respect, on December 25th 1968, Yourgrau wrote to Popper: "PLEASE, write me something about quantum theory." PA 106/9

(Popper 1971), which was probably one of the first reactions to the (at that time) overlooked and yet pivotal results of Bell's inequality (Bell 1964) and its experimentally testable form (Clauser et al. 1969).

Together with Henry Margenau, Yourgrau founded, in 1969, the journal *Foundations of Physics*, in whose pages physicists finally found new room for foundational speculations. This endeavor also involved Popper from the very beginning (see Section 4).

In conclusion, the unconditional support of Bunge and Yourgrau helped a great deal to bridge Popper's activity on the FQM and the *Denkkollektiv* of physicists.

### *3.2 Quantum Mechanics without the Observer*

*Quantum Mechanics without the Observer* has been referred to as Popper's "most influential essay on the topic" of FQM (Howard 2004), and it is out of doubt that this marked Popper's entry into the *Denkkollectiv* of physicists.

This publication was, perhaps for the first time (from the 1930s), clearly intended for an audience of physicists, and Popper himself points this out, although in a jokingly way: "I am sure I will shock many physicists who […] will stop reading this rubbish" (Popper 1967, p. 15). Given the scope of the paper, I deem it important to summarize the salient points thereof (some of Popper's arguments have been commented in detail in e.g. Jammer 1974; Jammer 1991 and Howard 2012).

Already in the first lines, Popper states that "this is an attempt to exorcise the ghost called 'consciousness' or 'the observer' from quantum mechanics" (Popper 1967, p. 7). In fact, although it is well known that the CIQM has not a definite and unique formulation (see Section 1, footnote 1), Popper's attack was directed against Bohr's and Heisenberg's epistemologies, in so far as both of them (yet with differences) admittedly advocated the impossibility of observing a system without disturbing it, and thus the necessity of "measuring agencies".

Popper then continues pointing out that a major flaw in Copenhagen's argument was to confound the alleged role of a conscious observer, with the role that the scientist has in consciously providing models and theories, on the basis of which experiments are devised:

> our theories which guide us in setting up our experiment have of course always been our inventions: they are inventions or products of our 'consciousness'. But that has nothing to do with the scientific status of our theories which depend […] on their truth (correspondence to reality), or nearness to truth. (Popper 1967, p. 11)

Moreover, for Popper, the reason for such a confusion must be sought in the despicable instrumentalist description of theories, eventually leading back the problem to that of the aim of science:

> The view that theories are nothing but instruments, or calculating devices, has become fashionable among quantum theorists, owing to the Copenhagen doctrine that quantum theory is *intrinsically ununderstandable* […]. What is of real importance for the pure scientist is the *theory*. (Popper 1967, p. 12)

These are the main motivations for Popper's critique of CIQM, which were already present in his early works (Popper 1934). It is quite clear that since the aim of science is for Popper to formulate theories which gradually approximate truth (i.e. the correspondence to objective reality) any subjective role played by the "observer" would be incompatible or at least severely limited. Popper maintains that it is just a misconception to attribute the probabilistic character of quantum mechanics to a lack of knowledge, and this precisely "led to the *intrusion of the observer*" (Popper 1967, p. 17). Such a misconception is what Popper calls "*the great quantum muddle*" (Popper 1967, p. 18), namely, to treat statistical measure functions (proper of sample spaces) as physical properties of the element of the space. Moreover, as it had been since as early as 1934 and for the rest of Popper's career, also in *Quantum Mechanics without the Observer*, the main target of criticism is Heisenberg's uncertainty principle (see Del Santo 2017). Popper maintains that uncertainty relations are formally valid (they are a proven mathematical theorem), but their usual interpretation, which sets an epistemological limit to precision in measurement, is the result of a subjective interpretation of probability statements. Popper suggested in 1934 to interpret the uncertainty relations in a statistical fashion, that is, as referring to a statistical ensemble of particles. In such a way, he claims, when the value of a variable $q$ is known with precision $\Delta q$, it is only our predicted probability functions for a measurement of its conjugate variable $p$ that gets smeared by $\Delta p$ (according to the Heisenberg relation), but not the actual value of the measurement itself. Such a statistical interpretation based on ensemble of particles appears quite obsolete, in so far as it fails to consider predictions about single particles. In fact, one of the great novelties of *Quantum Mechanics without the Observer* is that here Popper, with his propensity interpretation, while keeping objective reality at the center of his agenda, provides a conceptual framework to treat single-case probabilities (see below).

A major achievement of Popper's 1967 essay is that he finally develops the connection between physical problems and his new interpretation of probability, as he had announced in his early publications on the propensity interpretation (Popper 1957; 1959). In fact, he stresses again that "the interpretation of the formalism of quantum mechanics is closely related to the interpretation of the calculus of probability" (Popper 1967, p. 28). The propensity interpretation elevates probability to a real physical property of the single experiment, or rather, of the experimental conditions. In this sense Popper takes inspiration from the "classical" interpretation of probability –which he attributes to Laplace– that assigns probabilities on the basis of fundamental (ontological) properties of systems. According to the classical interpretation, a six-face dice made of homogenous material has 1/6 probability to land on any given face. However, the classical interpretation seems to have considered only symmetry as an objective property that determines probabilities of events. On the other hand, the largely accepted frequency interpretation (e.g. in von Mises 1928) –which was uphold by Popper up until early 1950s– allows to objectively assign probabilities to events defined on ensembles. The necessity of finding the appropriate reference class and assigning probabilities as (virtually infinite) series of favorable instances over the total cases (frequencies) does not allow to unambiguously define probabilities for single events (i.e. events bound to happen only once, and that do not have a non-trivial reference class). Popper's propensity interpretation, instead, finds its origins in the classical interpretation, but it generalizes the concept to asymmetrical cases assigning 'weights' to possible outcomes, yet grounding these on properties of the whole "experiment" (that is of the system and its environment). This can be seen with the example of a loaded dice (for instance with a piece of lead inside, close to one of the faces) which affects the probability of the dice to land on a particular face. In such a way, probabilities are objective real properties of single events (as opposed to series of events) that are completely grounded on the physical properties both of the system in question and on everything that can influence the experiment (such as the gravitational field in the case of a dice). Namely, propensities are determined by all the factors in isolations that are considered influential to the experiment, and more in general by anything that lies within the past light-cone of the considered experiment. Thus, probabilities are so interpreted as propensities or tendencies to produce frequencies, that can however be assigned *a priori* (i.e. prior to any measured quantities) according to certain objective properties, and then can be experimentally tested by collecting the actual statistical frequencies.[24] In such a way even an event that cannot be replicated has a certain objective probability to occur (single-case probability). It ought to be stressed, however, that propensities do not have any natural relation with the probabilities entailed by quantum theory (i.e. given by the Born rule). Therefore, Popper, more or less explicitly, is forced to posit the existence of "quantum propensities", such that they show a wave-like behavior (i.e. their amplitudes add up) inducing the well-known interference patterns, for instance, in double-slit experiment.

Popper then intends to show that "the famous 'reduction of the wave packet […] is not an effect characteristic of quantum theory but of probability theory in general" (Popper 1967, p. 34). In fact,

---

[24] The relation between interpretations of probability, in particular Popper's propensities, and quantum physics has been presented in Del Santo, F., "The physical motivations for a propensity interpretation of probability, and the reactions of the community of quantum physicists" at the Symposium "Popper and the Philosophy of Mathematics" in Klagenfurt (Austria) on April 5[th]-7[th], 2018. Preliminary proceedings are available at https://www.aau.at/wp-content/uploads/2018/06/KPF_NL-4_1_Proceedings_final.pdf.

Popper illustrates with some examples (pinboard, semi-transparent mirror with single photon and toss of a coin) that one should consider different conditional probability distributions before and after a "measurement" takes place, and consequently the "reduction of the wave packet" is a trivial feature of probability theory. If one tosses a (fair) penny, for example, the probability of obtaining a certain outcome $a$ $(-a)$, say head (tail), out of the two possible ones, given the experimental arrangement $b$ (height of the toss, initial momentum, etc.) is *p(a,b)=1/2= p(−a,b)*. And such it remains until one does not look at the penny, but "if we bend down and look, it suddenly 'changes': one probability becomes 1 and the other 0". But we are now dealing with a different experiment (i.e. measuring the outcome, given the toss has happened already), and the associated probability distribution, in the case that the penny has landed head, would be *p(a,a)=1* and *p(−a,a)=0*. Popper concludes that

> there is no more involved here, or in any reduction of the wave packet, than the trivial principle: if our information contains the result of an experiment, then the probability of this result, relative to this information (regarded as part of the experiment's specification), will always trivially be p(a,a)=1.This also explains what is valid in von Neumann's principle […] that if we repeat a measurement at once, than the result will be the same with certainty. Indeed, it is quite trite that if we look at our penny a second time, it will still lie as before. (Popper 1967, p. 37)

Popper is here dealing with part of the so-called *quantum measurement problem*, and in particular with what is sometimes referred to as the "small" measurement problem.[25] This problem addresses the question: "why a certain outcome – as opposed to its alternatives – occurs in a particular run of an experiment?" (Brukner 2017). In fact, "not only quantum mechanics, but every probabilistic theory in which probabilities are taken to be irreducible 'must have' the small measurement problem" (Brukner 2017). But since Popper was a staunch supporter of irreducible indeterminism, clearly, he reduced the small measurement problem to a pseudo-problem, where only the tendency (propensity) of an event to occur is predefined.

Popper's partial trivialization of the collapse of the wave function fails in solving the second part of the measurement problem, the "big measurement problem", which addresses the question: "what makes a measurement a measurement?" (Brukner 2017. See also footnote 25). Following Popper's example, in fact, it is not clear what does "bend down and look" is supposed to represent in quantum theory. Clearly, the analogous of the quantum measurement in the classical example of Popper is the coin toss and not the action of bending down and looking, which only represents an update of knowledge of an already determined state. On the contrary Popper seems to maintain that both coin tossing and bending down and looking both represent measurements. Whereas in quantum physics, repeated actual measurements (in the same basis) yield to the same results. Such a difference is not emphasized enough by Popper. Thus, Popper's program of taking "the mystery out of the quantum theory" (Popper 1957), does not seem to have been satisfactorily accomplished, insofar as he had to posit that quantum probabilities are not reducible to classical ones and their "mystery" remains unexplained. Admittedly, Popper acknowledges, contrarily to what Feyerabend was then to allege (see Section 3.4), an inherent difference between classical and quantum physics, and as such, between classical and quantum probabilities. In fact, whilst concepts the likes of "reduction of the wave packet" are scaled down by Popper thanks to his propensity interpretation of probability, he is forced to admit that "the peculiarity of quantum theory is the principle of superposition of wave amplitudes – a kind of *probabilistic dependence* […] that has apparently no parallel in classical probability theory" (Popper 1967, p. 40).

Nevertheless, *Quantum Mechanics without the Observer* surely provided, at the time when it appeared, a genuine insight into some issues of FQM. Within its pages Popper proposed an innovative realistic interpretation and he cogently established an unprecedented link between interpretations of probability and interpretations of quantum mechanics.

---

25 This categorization into two distinct measurement problems was firstly introduced by J. Bub and I. Pitowsky (See Brukner 2017 and references therein). The second, the "big" measurement problem instead deals with providing an explanation to the sudden transition between the deterministic, unitary evolution of the quantum state (through the Schrödinger equation) and the irreversible, indeterministic measurement.

### *3.3 First Reactions of the physicists*

Immediately after the publication of the paper (1967), Popper garnered the support and appreciation of some illustrious physicists, surely also thanks to his already established personal acquaintances and friendships. Among them, Bohm was the first to write to Popper on March, 3$^{rd}$ 1967:

> I feel that what you have to say about propensities makes a genuine contribution to clarifying the issues that you discuss. (PA 84/19)

It seems that even Bohm was unaware of Popper's propensity interpretation, although Popper devised and published papers on it more than a decade before. Also, Bohm participated to the Bristol symposium of 1957, where Popper's propensity interpretation was presented for the first time (see Section 2, especially footnote 7) and had regular discussions with Popper. This confirms that Popper's propensity interpretation was completely ignored by physicists, even by the closest to him, but this surely happened also because – as I have shown – it was only with this paper (Popper 1967) that Popper succeeded in consistently relating the (quantum) physical motivations for propensities with the interpretation of formal probability.

On September 5$^{th}$, 1967, also Landé sent to Popper a letter of appreciation for his recent work:

> thank you for sending me your brilliant paper about the 'Observer' which I read with the greatest pleasure and satisfaction. […] I admire your patience with showing again the nonsense of the contracting wave packet. But people who once have received the Nobel Prize turn out to be incorrible (*sic*) and unable to see through the 'great quantum muddle' of their own creation (letter of Landé to Popper, PA 319/18).

Bondi reacted positively to Popper's paper, too. He had been giving lectures on QM at the King's College of London for at least the two-year period 1965-67, immediately prior to Popper's publication. Interestingly, Bondi directly acknowledged Popper for the influence that the latter's ideas had on his course on QM, and expressed his overall agreement with Popper's critical paper:

> I have been reading your 'Quantum Mechanics without the Observer' which you sent me, with great interest and enjoyment. Some of it was quite new to me. […] The eye-opener for my treatment of this course was your remark two years ago that quantum theory gives statistical answers because one asks statistical questions. On the basis of this extremely penetrating remark I had myself come to the conclusion that the notion of the Observer was redundant". (Bondi to Popper, on September 11$^{th}$, 1967. PA 278/9).

What is however truly remarkable, is that in the following months the consensus about Popper's positions on QM enlarged to include new influential physicists, outside of Popper's circle of acquaintanceship. In 1968, the Swiss mathematician Paul Bernays, showed the paper *Quantum Mechanics without the Observer* to his colleague Bartel L. van der Waerden. The latter was a pupil of Emmy Noether in Göttingen, and developed group theoretical methods for QM, then becoming interested in FQM in Leipzig, probably thanks to his close collaboration with Heisenberg. On October 19$^{th}$, 1968, van der Waerden wrote to Popper some flattering remarks concerning his recent paper:

> I fully agree with your 13 theses, and I feel it was very good you expounded them so clearly. I also agree with your propensity interpretation of probability. […] I am sending you a paper on Measurements in Quantum Mechanics, in which I too have discussed the role of the Observer. I feel my ideas are in perfect accordance with your theses" (letter from van der Waerden to Popper. PA 96/27).

Indeed, van der Waerden had just published a paper (van der Waerden 1966) in which he claimed that the reduction of the wave packet or any "quantum jumps" are not a necessary feature of quantum physics, although they give correct predictions.

These words of appreciation surely must have been fresh air to Popper, who was finally getting the consideration that he had been waiting for decades. He enthusiastically answered:

I hardly need to say that I agree with every word of [your paper]. Your paper – and even more your letter – mean more to me than you can possibly imagine. It would take a long letter to give you an idea of my very lonely 35 years struggle. Although I had some encouragement, there was much more that was discouraging; and your letter is by far the most powerful encouragement I ever received." (Letter from Popper to van der Waerden on 28/10/1968. PA 96/27).

Meanwhile, Wolfgang Yourgrau, also read Popper's recent papers on QM and he brought it to the attention of the French Nobel laureate and founding father of quantum theory, Louis de Broglie. The latter wrote a short but appreciative note to Popper: "Yourgrau has sent me two of your articles on the interpretation of Quantum Mechanics. I noticed with great pleasure that your ideas are very close to mine".[26] (March 4th, 1969. PA 96/7).

In the following years, Popper's paper became known to some other physicists, like Leslie Ballentine who largely quoted and appreciated Popper's work (e.g. in Ballentine 1970) and embraced his statistical interpretation, and eventually to Vigier and Franco Selleri (see Del Santo 2017).

Popper's paper received also critical responses: Jeffrey Bub, a former student of Bohm in London, rebutted Popper's propensities (Bub 1972), formally developing the criticisms levelled by Feyerabend in his papers of 1968-69 (see Section 3.4 and Jammer 1974, pp. 452-453). Bub shows in detail that the main issue with Popper's interpretation of QM in terms of propensities is that Popper provides an interpretation (alternative to the frequency one) of a Boolean probability calculus which is not compatible with quantum probability. This critique is sound, and it formalizes the aforementioned problem of the addition of amplitudes in quantum theory but does not completely reflect Popper's views. Although Popper never formalized these concepts, he explicitly acknowledges that the probabilistic dependence of QM "has apparently no parallel in classical probability theory" (see Section 3.2).[27]

### 3.4 The controversy with Feyerabend over Quantum Physics

Another Viennese eminent philosopher of science of the last century, who also contributed much to the FQM, was Paul K. Feyerabend (1924-1994). Feyerabend studied physics at the University of Vienna, but he then turned to philosophy, completing his doctoral thesis in 1951 under the supervision of Viktor Kraft, at that time the last Vienna Circle's survivor in Vienna. After his graduation, Feyerabend moved to London to study the philosophy of QM under Popper's supervision. However, in 1953, he refused to become Popper's research assistant and moved back to Vienna where he published, in 1954, his first paper on the philosophy of QM, entitled *Determinismus und Quantenmechanik* (Feyerabend 1954). Therein, the author firmly advocated indeterminism, even in classical physics. Such an anti-deterministic position was obviously strongly influenced by Popper's (see Sect. 2), as Feyerabend explicitly points out in his conclusions. In the same paper, Feyerabend discusses von Neumann's impossibility proof for 'hidden variables' in quantum mechanics.[28] This became for Feyerabend the main object of investigation in the following years and, in April 1956, he reviewed his argument against von Neumann's impossibility proof submitting a short paper (Feyerabend 1956) to the prestigious German physics journal *Zeitschrift für Physik*. Therein, the author demonstrated that the logical flaw in von Neumann's argument is that it applies also to classical (statistical) theories and it is not inherent in QM, having nothing to do with the existence of 'hidden variables'. When informed of this publication to come (by Feyerabend himself on April 14th, 1956. PA 537/1) Popper was upset since he basically accused Feyerabend of plagiarism. He wrote to his former student some harsh letters,[29] whose contents deserve to be extensively reproduced:

> I have read your new paper. I do agree of course with its contents, since you got the complete contents of this paper here, in my room, from Joske [Joseph Agassi] and myself. […] I may remind you again of the fact that, when you still believed in the Neumann Proof you were here, and we had a great fight about it. […] If you do not add a very full acknowledgement to Joske and myself, it will [be] plagiarism. (Draft of letter from Popper to Feyerabend on April 16th, 1956. PA 537/1).

---

[26] "Yourgrau m'a communiqué deux articles de vous sur l'interprétation de la Mécanique Quantique. J'ai constaté avec grand plaisir que vos idées se rapprochent beaucoup des miennes.". The second paper that de Broglie referred to in the letter is almost surely (Popper 1968).

[27] Many physicists have subscribed to some variants of propensity interpretation, as a way to maintain both indeterminism and realism. In particular Nicolas Gisin has provided a non-Boolean calculus of probability (i.e. an axiomatization that does not satisfy Kolmogorov axioms) which provides a mathematical framework for propensities (Gisin 1991). This could possibly be seen as a rebut of Bub's argument.

[28] Historically, the main line of research for restoring realism in QM is to conceive a set of 'hidden variables' that complete the quantum state to encompass the "real state of affairs", i.e. the values of all the physical properties independently of any measurements. In 1932, Von Neumann allegedly proved that no hidden variable theory could be consistent with QM. This was however questioned and eventually disproved by Bohm, J. S. Bell, S. Kochen and E. Specker among others. For a historical, critical reconstruction see e.g. (Dieks 2017).

And Popper reinforced his message, a few days later:

> I do not believe that the contents of your paper are really worth publishing. I should have never published it myself. It is thus not that I think you are taking <u>valuable</u> goods from Joske and myself – goods which <u>we</u> find valuable. It is a different thing: a matter of principle. The goods are goods which <u>you</u> find valuable enough to publish. And they are not your property. […] I shall express my view that I consider a publication to be morally wrong. […] I should have willingly agreed to have it published by you, provided you would have proceeded in the proper way. […] I have to do it for your sake: otherwise you will never learn what one may do and not do (Draft of letter from Popper to Feyerabend on April 18th, 1956. PA 537/1).

After apologizing, Feyerabend accepted to include in the proofs an endnote written by Popper.[30] Yet, "he eventually prevented the note from appearing, confessing to an inquiring Agassi that he thought that the acknowledgement which Karl [Popper] had dictated was much too strong, but [he] ha[d] been afraid of quarrel" (Collodel 2016). This unfortunate episode did not really undermine the relationship between Popper and Feyerabend, who for some more years remained aligned with Popper's realist standpoint and in open opposition with CIQM.

However, after mid-1960s Feyerabend matured an increasingly more severe and structured critique of Popper's views on scientific method as well as on quantum physics. John Preston, in his thorough book on Feyerabend, notices that "there is something obsessive about the way in which Popper became Feyerabend's favourite whipping-post" (Preston 1997, p. 212, footnote 3). Accordingly, after Popper had published his *Quantum Mechanics without the Observer*, Feyerabend immediately reacted publishing, in December 1968,[31] the first part of the paper "On a Recent Critique of Complementarity" (Feyerabend 1968), wherein the author harshly criticized his former mentor, taking instead side in favor of Bohr, and of CIQM along with it:

> The publication of Bunge's Quantum Theory and Reality and especially of Popper's contribution to it are taken as an occasion for the restatement of Bohr's position and for the refutation of some quite popular, but surprisingly naive and uninformed objections against it. Bohr's position is distinguished both from the position of Heisenberg and from the vulgarized versions which have become part of the so-called 'Copenhagen Interpretation' and whose inarticulateness has been a boon for all those critics who prefer easy victories to a rational debate" (Feyerabend 1968).

Feyerabend levels a fair criticism when he affirms that it is imperative to make the necessary distinctions between Bohr, Heisenberg and the so-called CIQM: a distinction that in fact Popper seldom considered. I shall briefly analyze the core argument of Feyerabend's criticism. This is mainly based on a misconception of Popper's aim in proposing the pinboard argument (see Section 3.2). Indeed, Feyerabend writes:

---

[29] While finalizing the present paper I became aware of the imminent publication of a volume collecting all the survived letters between Popper and Feyerabend, edited by Matteo Collodel and Erik Oberheim. The unpublished letters here quoted will thus be soon accessible in transcript in a systematic collection.

[30] A draft of this note is attached to the already quoted letter from Popper to Feyerabend (on April 16th, 1956. PA 537/1), it begins with: "Diese Mitteilung ist wesentlichen ein Resultat von Diskussionen mit Professor K.R. Popper (London) und seinem Assistenten, Dr. J. Agassi." ("This communication is essentially a result from discussions with Professor K.R. Popper (London) and his assistant, Dr. J. Agassi.").

[31] A first version of the manuscript was rejected by *The British Journal for Philosophy of Science* in late 1967, and Feyerabend managed to have it published in *Philosophy of Science*, after a revision which incorporated replies to the comments made by the journal referee and by Bohm's pupil Jeffrey Bub (see Collodel 2016). Bub then developed Feyerabend's criticisms against Popper's propensities (see Section 3.3).

> [Popper] pleads with us not to be surprised when a change of experimental conditions also changes the probabilities. […] Quite correct-but irrelevant. For what surprises us (and what led to the Copenhagen Interpretation) is not the fact that there is some change; what surprises us is the kind of change encountered: trajectories which from a classical standpoint are perfectly feasible are suddenly forbidden and are not entered by any particle.

What Feyerabend means in this passage is that the novelty of quantum physics relies on the evidence that the quantities added are now the amplitudes of the wave function –whose norm square gives the associated probabilities–and not directly the probabilities; this gives raise to the genuine quantum behavior of interference. This is surely an important point, but not completely pertinent to Popper's argument, since Popper clearly stresses that the pinboard example has a different aim than that of trivializing the problems of probability in QM. His aim is to show that different experimental arrangements give raise to different probability distributions, even for single events and in the familiar classical physics, thus legitimizing the propensity interpretation which was explicitly devised to be applicable to single events. As I have showed, however (Section 3.2), Popper does not manage to solve the fundamental issues of quantum mechanics using propensities, for their composition has to be posited to recover the predictions of quantum mechanics (interference terms).

Bartley vindicated Popper against Feyerabend's attacks in his essay "The Philosophy of Karl Popper: Consciousness and Physics" (Bartley 1978). Therein he severely criticized Feyerabend's rebuttal of Popper's pinboard argument, rightly noticing that Popper never intended to explain interference pattern with the pinboard argument, as Feyerabend allegedly claimed. Bartley concludes that "Feyerabend's mistake here is extraordinary" and "that Feyerabend's objection to Popper's analysis of this issue […] is without merit." (Bartley 1978, p. 695).

There is yet another interesting historical hypothesis to be pursued. In fact, it is in my opinion quite clear that Feyerabend also used the ground of foundations of QM to openly attack Popper, in a much vaster campaign conducted in those years against Popper's ideas (especially on his epistemological views). An unpublished letter from Feyerabend is particularly significant to shedding new light on the motivation that led him to level this devastating criticism against Popper:

> We have here [at University of Berkeley, California] a few students of physics who are thoroughly critical of the way in which quantum theory is being taught today and who just managed to introduce an official course, run by them, in which they want to explore the weakness of the orthodox interpretation. I gave them copies of your article [(Popper, 1967)] and I never saw such an enthusiastic response. […] They also asked me for my opinion (they were all in my course on philosophy of science). As a reply I wrote the enclosed note which I also sent to BJPS to be published as a <u>discussion note</u> there. On rereading it I find that on various occasions I have expressed myself rather harshly […] but I don't think this will do any harm (Feyerabend to Popper on October 4th, 1967. PA 295/8).[32]

Feyerabend's critique is not devoid of personal resentment, as he also makes evident in a letter written to Popper's colleague John W. Watkins, immediately after the publication of the paper: "one of the reasons why I was mad at Popper was that his paper [(Popper, 1967)] did not pay any attention to my criticism of 1962 [(Feyerabend 1962)]. Maybe he had not read my paper (which I sent him); maybe he did not like it." (Feyerabend to Watkins, December 17th, 1967, reported in Collodel 2016).

Feyerabend's critical paper reached the *Denkkollektiv* of physicists, too. Margenau was personally quoted in Feyerabend's paper, and wrote him a vitriolic letter:

> I am amazed that you should feel called to defend a physicist like Bohr, whose work is understood by all. Those of us who think little of his complementarity principle do not underestimate its original heuristic force, but discard it because recent developments in which he did not participate have now made it pointless. And your archaic examples and superficial verbalisations, far from changing that situation, greatly emphasize it. […] I shall now comment briefly on matters that pertain to me in so far as they have been either misunderstood or vilified in your article. […] Your quotation on page 316 accuses me of a stupidity which working physicists resent. […] Whether you accept my distinction between possessed and latent or dormant observable is unimportant; men whose judgement I respect have thought it significant. But to brush it off as fancy terminology without probing the substance and, worse, to indict my students along with me in this context is less than generous. It is indeed a gesture I have not previously encountered in my long career at Yale.

---
[32] Berkley was to become within a few years the cradle of a counterculture against scientific orthodoxy, which found its expression in a New Age, hippie mood. See (Kaiser, 2011).

Popper's point of view, against which you argue, is at least philosophically interesting, even though physicists have paid little attention to it. I do not see that you have demolished it. […] You have stimulated me to say these things more formally in print." (Margenau to Feyerabend on February 17th, 1969. PA 96/12).

Popper answered Margenau's letter, recalling some unpleasant episodes, of the kind I have described above:

Feyerabend was once a student of mine, and I treated him extremely well. He never got over it: there are people who can never forget a benefit received, and other who cannot forgive it. Moreover, Feyerabend has no original ideas, but he poses as an original thinker: he is a compulsive plagiarist. He has stolen many of my ideas and although sometimes admitted this, he continues to criticise me, using my own ideas in his criticisms (Popper to Margenau on February 21st, 1969. PA 324/5)

Also Landé showed his solidarity, in a letter addressed to Popper on June 1st, 1969, and after he had received the second part of Feyerabend's paper (Feyerabend 1969) in defense of Bohr's complementarity (published on March 1969), he commented:

it contains a number of quite intemperate (and entirely mistaken) attacks against you and Bunge, the main argument being that you 'did not understand Bohr'. I intend to write a short rebuttal, less of Feyerabend than of the Copenhagen superstition from the purely physical point of view […]." (Landé to Popper. PA 318/18)

Popper replied a few days later with a somehow peculiar resignation, considering his usual combative temperament. He wrote to Landé:

<u>I heard about it</u>. I do not want to read it because why should I get angry? (Feyerabend is of course one of my former students for whom I did more than any teacher can be expected to do and has behaved to me in return simply disgustingly.) I should be most grateful for your defense if you would defend me. […] However, if you do not want to say so, simply don't: I certainly have no claim to be defended by anybody else, and I do not wish to defend myself because this would mean reading this silly stuff. (Popper to Landé. PA 318/18)

In the end, however, neither Landé nor Margenau, and certainly not Popper, publicly rebutted Feyerabend's theses.

### 4. Aftermath

As a consequence of the many critical activities that have been here expounded, the years after 1967-68 have been characterized by a certain public acknowledgement of Popper's belonging in the *Denkkollektiv* of quantum physicists. For example, in April 1968, Popper was invited as a speaker to the meeting "Quantum Theory and Beyond" at the King's College in Cambridge.[33] The meeting was completely organized by preeminent physicists, having Ted Bastin and David Bohm as proposers and Leon Rosenfeld as sponsor. Although Popper could eventually not take part in the colloquium, this meeting shows the consideration that Popper's opinion had at that time in the small community concerned with fundamental issues in QM. In fact, this time the participants were only physicists (among whom stand out the names of Bohm, O. R. Frisch, C. F. von Weizsäcker; G. M. Prosperi; J. Bub, Y. Aharonov, R. Penrose, B. J. Hiley, R. H. Atkin, T. Bastin, M. Bunge, H. R. Post). It is also noteworthy that Popper's effort to reinterpret Heisenberg uncertainty relations is mentioned already in the introduction of the proceedings together with Einstein's (Bastin 1971, pp. 5, 8).

---

[33] The invitation letter, signed by Bastin, Bohm and Rosenfeld, stated "There seem to be a number of possible directions of development of the subject [quantum theory] open at present. We aim to represent all of these at the meeting […]" (Letter to Popper on April 25th, 1968; PA 278/2). It is thus remarkable that they consider Popper's line of research on QM, as one of the possible directions of developments of the theory.

In 1968, Popper published in the very prestigious journal *Nature* a controversial paper wherein he claimed to have found a crucial mistake in a famous work of G. Birkhoff and von Neumann that inaugurated the so-called Quantum Logic. This led Popper to have heated discussions with physicists pursuing that approach like J. M. Jauch, D. Finkelstein, A. Ramsay and J. Pool (see Section 1, footnote 3, and Jammer 1974, pp. 353 ff.). Popper also served as referee for *Nature* in several occasions after that episode, between 1968 and 1971 (PA 331/10).

Moreover, in January 1969, Yourgrau and Margenau founded the new journal *Foundations of Physics* and Popper immediately joined the Editorial Board, together with distinguished physicists such as P. Bergmann, L. de Broglie, M. Gell-Mann and E. Wigner. Popper also served as referee for that journal, in 1972 and in 1973 for papers on FQM (PA 296/22).

Also, important physical institutions started considering Popper as a valuable source for foundational papers specifically aimed at an audience of physicists. This is the case of the renowned *Institute of Physics and the Physical Society* which explicitly asked Popper to write "an authoritative article for Physics Bulletin about the current ideas of philosophy of the nature and basis of fundamental science." (Letter from Kurt Paulus on May 24[th], 1968. PA 311/25).

At the University of London from the late sixties and throughout the seventies the fundamental issues at the edge between physics and philosophy became more openly discussed (Bohm had been appointed a chair in theoretical physics at the Birkbeck College, in walking distance from the London School of Economics where Popper was professor). Basil Hiley, perhaps the closest collaborator of Bohm, recalls regular seminars on quantum theory, which he used to attend together with Bohm, Popper and Vigier visiting from Paris.[34]

We must stress that the Irish physicist John S. Bell, as early as 1964, had put forward a momentous theorem (Bell 1964) that is still at the basis of contemporary research on FQM, allowing to experimentally discriminate between local realistic hidden variable theories and QM. However, also because of the productivistic scientific programs of the post-war period – summarized by the expression "Shut up and calculate" – this striking result was prevented from having almost any resonance for several years. However, at the end of 1960s, the scientific atmosphere was slightly changing, and some pockets of resistance were to breathe new life in the FQM, recognizing the utmost relevance of Bell's inequalities (see Freire 2014 and references thereof). The room for speculation about realism was moving from the community of philosophers to that of physicists, and later on literally into the laboratories. Popper became aware of Bell's inequalities thanks to Abner Shimony – one of the initiators of the research on this topic (see Clauser *et al.* 1969) – when the latter sent him a copy of Bell's first paper on the subject, on November 30[th], 1969 (PA 350/7). Besides, Popper was one of the first to respond in print to the novel, striking findings – at that time still almost completely overlooked – of Bell and of Clauser, Horn, Shimony and Holt, discussing them extensively in his contribution to Landé's *Festschrift*, entitled *Particle Annihilation and the Argument of Einstein, Podolsky and Rosen* (Popper 1971).

At that time, the *Denkkollektiv* of quantum physicists was enlarging its horizons tremendously, and Popper was one of the few elements of continuity during this change. In fact, thanks to his intense participation to the debate over FQM in the 1960s and his fame as a philosopher of science, it was not hard for Popper to find room in the *Denkkollektiv* of physicists, even throughout this generational transition. He indeed established or strengthened long lasting and regular interactions with physicists the likes of Vigier, J. Clauser, A. Shimony, D. Deutsch, S. Kochen, E. P. Wigner, F. Selleri, H. Rauch, etc., whose ideas were about to have a prime influence in the new 'quantum era' to come (see again Freire 2014). In the 1980s, Popper's role in the *Denkkollektiv* became even more manifest and quantum entanglement and Bell's inequalities became Popper's main research topic in physics ever after (see Del Santo 2017).

---

[34] I am most grateful to Prof. Hiley for sharing this testimony in a personal communication (Basil Hiley, email to the author, October 10, 2016).

In conclusion, this paper shows that Popper deserves to be considered a fully-fledged "quantum dissident", namely a member of a *Denkkollektiv* of physicists that questioned FQM at a time when physics had lost its philosophical roots.

**Aknowledgements**

I am thankful to o Mag.[a] Nicole Sager and Dr. Manfred Lube from the *Karl Popper Sammlung* of the AAU in Klagenfurt for granting me the access to their archives and the reproduction of Popper's original correspondence.

I wish also to express my gratitude to Prof. Bernard Burgoyne for useful comments.

This research did not receive any specific grant from funding agencies in the public, commercial, or not-for profit sectors.

**References**

Ballentine, Leslie E. 1970. "The Statistical Interpretation of Quantum Mechanics." *Reviews of Modern Physics* 42, 358- 381.

Bartley, William III. 1978. "Critical Study: The Philosophy of Karl Popper, part II: Consciousness and physics." *Philosophia* 7, no. 3: 675-716.

Baracca, Angelo, Silvio Bergia, and Flavio Del Santo. 2016. "The Origins of the Research on the Foundations of Quantum Mechanics (and other Critical Activities) in Italy during The 1970s." *Studies in History and Philosophy of Modern Physics* 57: 66-79.

Bastin, Ted. 1971. *Quantum Theory and Beyond. Essays and Discussions Arising from a Colloquium.* Cambridge: Cambridge University Press.

Bell, J. S. 1964. "On the Einstein-Podolsky-Rosen paradox". *Physics*, 1, 195-200.

Bohm, David. 1952. "A Suggested Interpretation of the Quantum Theory in Terms of Hidden Variables I." *Physical Review* 85: 166-179.

Brukner, Č. 2017. "On the quantum measurement problem". in *Quantum [Un]Speakables II: Half a Century of Bell's Theorem*, Bertlmann, R. & Zeilinger, A., eds. 2017, Springer, pp. 95–117.

Bub, Jeffrey. 1972. "Popper's Propensity Interpretation of Probability and Quantum Mechanics." University of Minnesota Press, Minneapolis. Retrieved from the University of Minnesota Digital Conservancy. 1975. http://hdl.handle.net/11299/184671.

Bunge, Mario. 1955. "Strife about Complementarity." *The British Journal for the Philosophy of Science* 6, no. 22: 141-154.

Bunge, Mario, ed. 1964. *The Critical Approach to Science and Philosophy*. Ney York: The Free Press of Glencoe.

Collodel, Matteo. 2016. "Was Feyerabend a Popperian?". *Studies on History and Philosophy of Science Part A* 57: 27-56. doi: 10.1016/j.shpsa.2015.08.004

Clauser, John F., Michael A. Horne, Abner Shimony, Richard A. Holt. 1969. "Proposed Experiment to Test Local Hidden-Variable Theories." *Physical review letters* 23, no. 15: 880.

Del Santo, Flavio. 2017. "Genesis of Karl Popper's EPR-like Experiment and its Resonance amongst the Physics Community in the 1980s." *Studies in History and Philosophy of Modern* Physics, 62: 56-70. https://doi.org/10.1016/j.shpsb.2017.06.001.

Dieks, Dennis. 2017. "Von Neumann's Impossibility Proof: Mathematics in the Service of Rhetorics." *Studies in History and Philosophy of Modern Physics* 60: 136-148. https://doi.org/10.1016/j.shpsb.2017.01.008.

Feyerabend, Paul K. 1954. "Determinismus und Quantenmechanik." *Wiener Zeitschrift für Philosophie, Psychologie, Pädagogik* 5: 89-111. Reproduced in English translation in *Physics and Philosophy: Philosophical Papers*, Volume 4, Stefano Gattei and Joseph Agassi, eds. 2016. Cambridge: Cambridge University Press, pp. 25-45.

Feyerabend, Paul K. 1956. "Eine Bemerkung zum Neumannschen Beweis." *Zeitschrift für Physik* 145, no. 4: 421-423. Reproduced in English translation in *Physics and Philosophy: Philosophical Papers*, Volume 4, Stefano Gattei and Joseph Agassi, eds. 2016. Cambridge: Cambridge University Press, pp. 46-48.

Feyerabend, Paul K. 1962. "Problems of Microphysics." In R. G. Colodny, ed., *Frontiers of science and philosophy*, 189-283. Pittsburgh: University of Pittsburgh Press.

Feyerabend, Paul K. 1968. "On a Recent Critique of Complementarity: Part I." *Philosophy of Science* 35: 309–333.

Feyerabend, Paul K. 1969. "On a Recent Critique of Complementarity: Part II." *Philosophy of Science*, 36: 82-105.

Fleck, Ludwig. 1935. *Entstehung einer wissenschaftlichen Tatsache–Einführung in die Lehre vom Denkstil und Denkkollektiv*. Frankfurt: Main.

Freire, Olival Jr. 2014. *The Quantum Dissidents: Rebuilding the Foundations of Quantum Mechanics (1950-1990)*. Berlin: Springer.

Gisin, Nicolas. 1991. "Propensities in a non-deterministic physics". *Synthese*, 89(2): 287-297.

Hiley, Basil. 1997. "David Joseph Bohm. 20 December 1917-27 October 1992." *Biographical Memoirs of Fellows of the Royal Society* 43: 106- 131.

Howard, Don. 2004. "Who Invented the 'Copenhagen Interpretation'? A Study in Mythology." *Philosophy of Science* 71, no. 5: 669-682.

Howard, D. (2012). Popper and Bohr on realism in quantum mechanics. *Quanta*, 1(1): 33-57.

Jammer, Max. 1974. *The Philosophy of Quantum Mechanics*. New York: Wiley.

Jammer, Max. 1991. "Sir Karl Popper and His Philosophy of Physics." *Foundations of Physics* 21, no. 12: 1357-1368.

Kaiser, David. 2011. *How the Hippies Saved Physics: Science, Counterculture, and the Quantum Revival*. New York: WW Norton & Company.

Kožnjak, B. (2017). "The Missing History of Bohm's Hidden Variables Theory: The Ninth Symposium of the Colston Research Society, Bristol, 1957." *Studies in History and Philosophy of Modern Physics*, in press. https://doi.org/10.1016/j.shpsb.2017.06.003.

Kragh, Helge. 2013. "The most philosophically of all the sciences: Karl Popper and physical cosmology." *Perspectives on Science* 21, no. 3.

Landé Alfred. 1960. *From Dualism to Unity in Quantum Physics*. Cambridge: Cambridge University Press.

Mises von, R. 1928. *Wahrscheinlichkeit, Statistik und Wahrheit*. Vienna: Julius Springer.

PA, Popper's Archives, Box/Folder: 84/19. AAU, Klagenfurt (Austria)/Hoover Institution, Stanford (California).

PA, Popper's Archives, Box/Folder: 96/7. AAU, Klagenfurt (Austria)/Hoover Institution, Stanford (California).

PA, Popper's Archives, Box/Folder: 96/12. AAU, Klagenfurt (Austria)/Hoover Institution, Stanford (California).

PA, Popper's Archives, Box/Folder: 96/27. AAU, Klagenfurt (Austria)/Hoover Institution, Stanford (California).

PA, Popper's Archives, Box/Folder: 106/9. AAU, Klagenfurt (Austria)/Hoover Institution, Stanford (California).

PA, Popper's Archives, Box/Folder: 278/2. AAU, Klagenfurt (Austria)/Hoover Institution, Stanford (California).

PA, Popper's Archives, Box/Folder: 278/9. AAU, Klagenfurt (Austria)/Hoover Institution, Stanford (California).

PA, Popper's Archives, Box/Folder: 280/25. AAU, Klagenfurt (Austria)/Hoover Institution, Stanford (California).

PA, Popper's Archives, Box/Folder: 280/26. AAU, Klagenfurt (Austria)/Hoover Institution, Stanford (California).

PA, Popper's Archives, Box/Folder: 294/19. AAU, Klagenfurt (Austria)/Hoover Institution, Stanford (California).

PA, Popper's Archives, Box/Folder: 295/8. AAU, Klagenfurt (Austria)/Hoover Institution, Stanford (California).

PA, Popper's Archives, Box/Folder: 296/22. AAU, Klagenfurt (Austria)/Hoover Institution, Stanford (California).

PA, Popper's Archives, Box/Folder: 311/25. AAU, Klagenfurt (Austria)/Hoover Institution, Stanford (California).

PA, Popper's Archives, Box/Folder: 318/18. AAU, Klagenfurt (Austria)/Hoover Institution, Stanford (California).

PA, Popper's Archives, Box/Folder: 319/18. AAU, Klagenfurt (Austria)/Hoover Institution, Stanford (California).

PA, Popper's Archives, Box/Folder: 324/5. AAU, Klagenfurt (Austria)/Hoover Institution, Stanford (California).

PA, Popper's Archives, Box/Folder: 331/10. AAU, Klagenfurt (Austria)/Hoover Institution, Stanford (California).

PA, Popper's Archives, Box/Folder: 350/7. AAU, Klagenfurt (Austria)/Hoover Institution, Stanford (California).

PA, Popper's Archives, Box/Folder: 530/10. AAU, Klagenfurt (Austria)/Hoover Institution, Stanford (California).

PA, Popper's Archives, Box/Folder: 537/1. AAU, Klagenfurt (Austria)/Hoover Institution, Stanford (California).

PA, Popper's Archives, Box/Folder: 548/8. AAU, Klagenfurt (Austria)/Hoover Institution, Stanford (California).

Peat, David. 1997. *Infinite Potential: The Life and Times of David Bohm*. Massachusetts: Addison Wesley.

Popper, Karl R. 1934. *Logik der Forschung. Zur Erkenntnistheorie der modernen Naturwissenschaft*. Vienna: Julius Springer.

Popper, Karl R. 1951. "Indeterminism in Quantum-Mechanics and in Classical Physics." *The British Journal for the Philosophy of Science* 1, no. 2: 117-133.

Popper, Karl R. 1957. "The Propensity Interpretation of the Calculus of Probability, and the Quantum Theory." In Körner, Stephan, ed. *Observation and Interpretation, Proceedings of the Ninth Symposium of the Colston Research Society, University of Bristol 1-4 April, 1957*.

Popper, Karl R. 1959. "The Propensity Interpretation of Probability." *The British Journal for the Philosophy of Science* 10, no. 37: 25-42

Popper, Karl R. 1967. "Quantum Mechanics without 'the Observer'." In Bunge, Mario, ed. *Quantum theory and reality*. Springer, Berlin, Heidelberg, pp. 7-44.

Popper, Karl R. 1968. "Birkhoff and von Neumann's Interpretation of Quantum Mechanics." *Nature* 219, no. 5155: 682-686.

Popper, Karl R. 1971. "Particle Annihilation and the Argument of Einstein, Podolsky and Rosen." In *Perspectives in Quantum Theory*, edited by Yourgrau, Wolfgang and Alwyn van der Merwe, 182-198. Cambridge (Massachusetts): MIT Press.

Perspectives in Quantum Theory (1971): 182-198.

Popper, Karl R. 1976. *Unended Quest: An Intellectual Autobiography*. London: Routledge

Popper, Karl R. 1982. Edited by William Bartley, III, Vol. III of the *Postscript to the Logic of Scientific Discovery: Quantum Theory and the Schism in Physics*. London: Hutchinson; Totowa: Rowman and Littlefield.

Preston, John. 1997. *Feyerabend: Philosophy, Science and Society*. Cambridge: Polity Press.

Schilpp, Paul A., ed. 1974. *The Philosophy of Karl Popper*. La Salle: Open Court.

van der Waerden, Bartel Leendert. 1966. "On measurements in quantum mechanics." *Zeitschrift für Physik* 190: 99-109.